# 12-photon entanglement and scalable scattershot boson sampling with optimal entangled-photon pairs from parametric down-conversion

Han-Sen Zhong[1,2], Yuan Li[1,2], Wei Li[1,2], Li-Chao Peng[1,2], Zu-En Su[1,2], Yi Hu[1,2], Yu-Ming He[1,2], Xing Ding[1,2], W.-J. Zhang[3], Hao Li[3], L. Zhang[3], Z. Wang[3], L.-X. You[3], Xi-Lin Wang[1,2], Xiao Jiang[1,2], Li Li[1,2], Yu-Ao Chen[1,2], Nai-Le Liu[1,2], Chao-Yang Lu[1,2], Jian-Wei Pan[1,2]

[1] Hefei National Laboratory for Physical Sciences at Microscale and Department of Modern Physics, University of Science and Technology of China, Hefei, Anhui, 230026, China

[2] CAS Centre for Excellence and Synergetic Innovation Centre in Quantum Information and Quantum Physics, University of Science and Technology of China, Hefei, Anhui 230026, China

[3] State Key Laboratory of Functional Materials for Informatics, Shanghai Institute of Micro system and Information Technology (SIMIT), Chinese Academy of Sciences, 865 Changning Road, Shanghai 200050, China

**Entangled photon sources with simultaneously near-unity heralding efficiency and indistinguishability are the fundamental elements for scalable photonic quantum technologies. We design and realize a degenerate entangled-photon source from an ultrafast pulsed laser pumped spontaneous parametric down-conversion (SPDC), which show simultaneously ~97% heralding efficiency and ~96% indistinguishability between independent single photons. Such a high-efficiency and frequency-uncorrelated SPDC source allows generation of the first 12-photon genuine entanglement with a state fidelity of $0.572 \pm 0.024$. We further demonstrate a blueprint of scalable scattershot boson sampling using 12 SPDC sources and a $12 \times 12$-modes interferometer for three-, four-, and five-boson sampling, which yields count rates more than four orders of magnitudes higher than all previous SPDC experiments. Our work immediately enables high-efficiency implementations of multiplexing, scattershot boson sampling, and heralded creation of remotely entangled photons, opening up a promising pathway to scalable photonic quantum technologies.**

Spontaneous parametric down-conversion (SPDC) [1] has been the most widely used workhorse for producing entangled-photon pairs, which was exploited for tests of Bell's inequalities [2–4], quantum key distribution [5–7], and dense coding [8]. The development of multi-photon interferometry [9], which relied on quantum interference between independent photons, opened the way to coherent control of a large number of photonic qubits. This allowed the generation of Greenberger-Horne-Zeilinger (GHZ) entanglement [10–17] and tests of GHZ theorem [18], and found many applications in quantum information protocols such as quantum teleportation [19–21], quantum metrology [22–24], quantum simulation [25] and boson sampling [26–30].

In the past two decades, the number of genuinely multi-particle entangled photons from SPDC has been increased up to ten [15,16]. Yet, a more scalable implementation remained challenging, largely to the lack of a *perfect* entangled-photon source produced by ultrafast laser pumped SPDC where the photonic entanglement fidelity, the collection efficiency, and the indistinguishability between independent photons are *simultaneously* engineered to close to unity. Such a perfect source can immediately enable previously formidable tasks, for example, scalable scattershot boson sampling [31], heralded two-photon entanglement at distant locations for Bell test and device-independent quantum key distribution [32–34], and serve as a scalable building block for multiplexing [35–37] that can overcome the probabilistic nature of SPDC.

In the SPDC [38], the conservation of energy and momentum can naturally induce strong correlations in multiple degrees of freedom between the two converted photons including their polarization, frequency and time. In the view of quantum engineering, the single photons should be efficiently prepared in a pure state with a single degree of freedom. However, usually the uncontrolled entanglement in the frequency and/or time can significantly degrade the entanglement in the polarization. In the early experiments, to eliminate the spectral correlation of the photon pairs, the most straightforward way was passive narrowband (typical linewidth ~3 nm) spectral filtering, which significantly sacrificed the brightness and collection efficiency of the entangled photons [10–13]. A more efficient method was to employ the interferometric Bell-state synthesizer [39] to separate the correlation between the polarization and the spectral bandwidth, allowing for a more selective (3-nm linewidth for *e*-polarized and 8-nm for *o*-polarized photons) and thus more economical narrowband filtering [14]. Further, beam-like SPDC [40] was developed with the photon pairs in the form of two separate Gaussian-like beams, which had higher brightness and efficiency coupling into a single spatial mode [15,41] than those from the non-collinear SPDC where the collection was at intersections of the two down-converted photon rings. We note that, however, in these previous multi-photon entanglement experiments [10–16], the *e*- and *o*-polarized photons were frequency-correlated, as evident from the observed tilted two-photon joint spectral intensity distribution, which caused a trade-off between the efficiency and the indistinguishability. There has been important progress in preparing heralded single photons in frequency-uncorrelated pure quantum state [42–44], yet the simultaneous combination of near-unity entanglement fidelity, indistinguishability, and collection efficiency remained an outstanding goal.

Here, we design and experimentally realize an optimal SPDC source of entangled photons at telecommunication wavelength by combining the techniques of frequency-uncorrelated and beam-like SPDC. For the photon pair free from any correlation in their spatiotemporal degrees of freedom, it is necessary that the two-photon joint amplitude function is factorable [45]. We find a suitable parameter regime that fulfills the condition using a BBO crystal with a thickness of 6.3 mm, pump laser wavelength of 775 nm (generating photon pairs centered around 1550 nm), and pulsed laser bandwidth of 5.5 nm [46]. Our design of entangled photon source in beam-like SPDC configuration is illustrated in Fig. 1a. The pulsed laser passes through an arrangement of two YVO$_4$ beam displacers (BDs) and half-wave plates (HWP) to separate the laser beam into two paths by 2.6 mm apart. The two beams focus on one β-barium-borate (BBO) crystal to generate two identical photon pairs in the states $|V_1\rangle|H_2\rangle$ and $|V_{1'}\rangle|H_{2'}\rangle$ via beam-like type-II SPDC, where the subscripts denotes the spatial modes as drawn in Fig. 1a. The $|V_{1'}\rangle|H_{2'}\rangle$ pair is then rotated using a HWP to its orthogonal state $|H_{1'}\rangle|V_{2'}\rangle$, and recombined with $|V_1\rangle|H_2\rangle$ into a single spatial mode using two BDs. Tilting the two BDs allows for precise temporal tuning and fine spatial compensation of the photon pairs that prepares them into an entangled Bell state: $(|H_1\rangle|V_2\rangle + |V_1\rangle|H_2\rangle)/\sqrt{2}$, with a measured visibility above 0.997 in the basis of $|\pm\rangle = (|H\rangle \pm |V\rangle)/\sqrt{2}$.

We use 30-nm bandpass filters to remove the small sidebands and measure the joint spectral intensity distribution shown in Fig. 1b, from which we extract a spectral purity of 0.99. Further, due to the group delay dispersion of the YVO$_4$ crystals (310 fs$^2$/mm) and the BBO crystals (79 fs$^2$/mm), we employ four pairs of dispersion-compensating prisms to eliminate the dispersion. To test the photon indistinguishability, we perform a Hong-Ou-Mandel quantum interference [47] between two independent SPDC photons. As shown in Fig. 1c, at zero delay the four-fold coincidence count shows a dip with a raw visibility of $0.962 \pm 0.011$, without using inefficient narrowband spectral filtering. Its slight deviation from the predicted visibility of 0.99 from Fig. 1b could be caused by residual dispersion of the pump laser.

Another important requirement is the simultaneously high heralding efficiency and brightness. Generally, due to conservation of momentum in SPDC, a lower momentum uncertainty of pump beam can lead to a higher collection efficiency. However, a larger pump beam waist could result in a lower pump energy density. Thus, there is a trade-off between the collection efficiency and brightness. We measure the heralding efficiency and brightness of the new SPDC source as a function of pump beam waist. As shown in Fig. 1d, at a pump beam waist of 260 μm, we obtain a two-photon count rate of 7100 Hz/mW and a heralding efficiency of 85%. At a pump beam waist of 1.9 mm, the brightness decreases to 170 Hz/mW whereas the heralding efficiency increases to 94%. Subtracting the channel loss in the optical path from the BBO to the single-mode fiber, we estimate a corrected heralding efficiency of 97% at 1.9-mm waist.

This high-performance entangled photon source immediately makes it possible to perform the first 12-photon entanglement experiment. By successively passing the laser through six BBO crystals (see Fig. 2), we first prepare six pairs of entangled photons. One photon from each pair is combined with the other five photons on a linear optical array of five polarization beam splitters (PBSs) that transmit *H* and reflect *V* polarization. Under this arrangement, post-selecting 12-photon coincidences implies that the output photons are either all *H* or *V* polarized—two cases are quantum mechanically indistinguishable—thus projecting them into a 12-photon Greenberger-Horne-Zeilinger (GHZ) state in the form of $(|H\rangle^{\otimes 12} + |V\rangle^{\otimes 12})/\sqrt{2}$.

To analyze the generated 12-photon state, we use a combination of wave plates and PBSs to measure the polarization of each individual photon, and use 24 superconducting nanowire single-photon detectors register the 12-channel coincidence counts. We use a suitable laser power of 1.8 W and a focal waist of 0.55 mm, where the detected 2-photon count rate is 2.0 MHz and the 12-photon coincidence is about one per hour. To validate the 12-photon entanglement, we first measure the 12-photon events in the *H/V* basis (see data in Fig. 3a) to calculate the population of $(|H\rangle\langle H|)^{\otimes 12} + (|V\rangle\langle V|)^{\otimes 12}$ over all the possible $2^{12}$ combinations. From Fig. 3a, we extract the population $P = 0.732 \pm 0.024$. We further measure all the photon in the basis of $(|H\rangle \pm e^{i\theta}|V\rangle)/\sqrt{2}$ to estimate the expectation value of the observable $M_\theta^{\otimes N} = (\cos\theta\sigma_x + \sin\theta\sigma_y)^{\otimes N}$, where $\theta = k\pi/12$, $k = 0, 1, 2, ..., 11$. The coherence of the GHZ state, determined by the two off-diagonal elements of its density matrix, can be calculated by $C = (1/12)\sum_{k=0}^{11}(-1)^k \langle M_{(k\pi/12)}^{\otimes 12}\rangle$. From Fig. 3b, we calculate $C = 0.419 \pm 0.041$. We can then determine the state fidelity of the generated 12-photon GHZ state by $(P+C)/2 = 0.576 \pm 0.024$, which exceeds the threshold 0.5 more than 3 standard deviations and is sufficient to prove the presence of a genuine 12-qubit entanglement [48].

Scaling up to a larger number of photons would be prohibited by the intrinsically probabilistic generation of entangled photon pairs in the SPDC, which can be overcome using multiplexing. For a scalable multiplexing with practical advantage over a single SPDC pair, it is crucial that the SPDC pair, which serves as the fundamental building block for multiplexing, should possess simultaneously near-unity heralding efficiency and photon indistinguishability as we have demonstrated here. Thus, by combining our SPDC sources with multiplexing with fast and low-loss switches and suitable optical memories [35–37], it is possible to significantly enhance the overall efficiency, opening a new path to large-scale linear optical quantum computing.

For applications in boson sampling [49]—a special model of quantum computing and considered as a strong candidate of "quantum computational supremacy"—there is a more convenient and efficient protocol to overcome the probabilistic problem of the SPDC without the need of multiplexing. The standard boson sampling is usually realized by sending $n$ indistinguishable single photons through an $m$-mode ($m > n$) interferometer, and registering the $n$-photon counts. Using $n$ heralded single photons as input from $n$ SPDC photon-pair sources, each with an generation probability of $\varepsilon$ per pulse and a heralding efficiency of $\eta$, the eventual $n$-photon count rate would scale as $(\varepsilon\eta)^n$. The key idea of scattershot boson sampling [31] is to use $k$ ($k \gg n$) heralded single-photon sources connecting to different input modes of the interferometer, which can achieve an exponential $\binom{k}{n}$ times increase in the $n$-photon count rate to compete against the intrinsic probabilistic loss $\sim \varepsilon^n$. A proof-of-principle demonstration of scattershot boson sampling has been reported previously [50], however, using inefficient SPDC sources with low $\eta$, which limited the scalability to larger number of photons.

We exploit the SPDC source with simultaneously near-unity indistinguishability and heralding efficiency to demonstrate a blueprint of scalable scattershot boson sampling. As shown in Fig. 4a, we use 12 SPDC sources to feed into a $12 \times 12$ mode interferometer encoded by both spatial and polarization degrees of freedom [46]. For each SPDC source, the idler photons are detected to herald the presence of the signal photons, which are combined into one path by two BDs and fed into the interferometer. There are 220, 495 and 792 different no-collision input combinations for the three-, four- and five-photon boson sampling, which implies that our scattershot boson sampling is expected to yield 220, 495 and 792 times enhancement of the efficiency over standard boson sampling, respectively.

We measure the three-, four- and five-photon sampling rate of 3.9 kHz, 44 Hz and 0.3 Hz, respectively. To qualify the sampling performance, we calculate the similarity, defined as $S = \sum \sqrt{p_i q_i}$, and the distance, defined as $D = (1/2) \sum |p_i - q_i|$, where $p_i$ and $q_i$ represent the experiment data and theoretical prediction, respectively. For the

3-boson sampling, averaging over the 220×220 input-output combinations, we obtain a similarity of 0.982±0.004 and a distance of 0.122±0.018. Such a characterization method is, however, not scalable to larger number of photons. For example, for the 4- and 5-boson sampling, the input-output combinations reach 495×495 and 792×792, respectively, such that the counting events in each combination are extremely scarce. Therefore, a more efficient method is adopted to validate the boson sampling data. We apply likelihood ratio test to rule out the distinguishable photon hypothesis [46]. The results are shown in Fig. 5b, with significant deviations between the experimental data (requiring only a few hundreds of samples) and the simulated distinguishable sampling.

The combination of our optimal SPDC source and the scattershot boson sampling yields a significantly enhanced multi-photon count rate. For example, the measured 3.9 kHz three-photon count rate is more than 4 orders of magnitudes higher than the best previous boson sampling experiments based on SPDC [26–29,50] and comparable to the work using state-of-the-art quantum-dot single photons (see Fig. 5c for a summary of the count rate of the 3-boson sampling experiments). We expect to further increase the efficiency by using more SPDC crystals, higher-efficiency detectors, and combining the protocol of photon-loss-tolerant boson sampling [51].

In summary, we have developed an optimal SPDC entangled-photon source with simultaneously near-unity indistinguishability and heralding efficiency, which allowed us to demonstrate the first 12-photon genuine entanglement and perform high-efficiency scattershot boson sampling. Our work has established an optimal multi-photon platform and will enable previously challenging experiments such as generations of high-NOON states and spatially separated two- and multi-photon entangled states with near-unity heralding efficiency, which will be useful for Bell tests with human's free will [33], long-distance device-independent quantum key distribution [32], quantum teleportation with remotely prior distributed entanglement [19–21], and demonstrations of quantum communication complexity [52]. Our SPDC source is also readily to be combined with multiplexing to overcome its probabilistic scaling, opening up a new pathway towards scalable photonic quantum technologies.

# References

[1]  P. G. Kwiat, K. Mattle, H. Weinfurter, A. Zeilinger, A. V. Sergienko, and Y. Shih, Phys. Rev. Lett. **75**, 4337 (1995).

[2]  G. Weihs, T. Jennewein, C. Simon, H. Weinfurter, and A. Zeilinger, Phys. Rev. Lett. **81**, 5039 (1998).

[3]  M. Giustina, M. A. M. Versteegh, S. Wengerowsky, J. Handsteiner, A. Hochrainer, K. Phelan, F. Steinlechner, J. Kofler, J.-Å. Larsson, C. Abellán, W. Amaya, V. Pruneri, M. W. Mitchell, J. Beyer, T. Gerrits, A. E. Lita, L. K. Shalm, S. W. Nam, T. Scheidl, R. Ursin, B. Wittmann, and A. Zeilinger, Phys. Rev. Lett. **115**, 250401 (2015).

[4]  L. K. Shalm, E. Meyer-Scott, B. G. Christensen, P. Bierhorst, M. A. Wayne, M. J. Stevens, T. Gerrits, S. Glancy, D. R. Hamel, M. S. Allman, K. J. Coakley, S. D. Dyer, C. Hodge, A. E. Lita, V. B. Verma, C. Lambrocco, E. Tortorici, A. L. Migdall, Y. Zhang, D. R. Kumor, W. H. Farr, F. Marsili, M. D. Shaw, J. A. Stern, C. Abellán, W. Amaya, V. Pruneri, T. Jennewein, M. W. Mitchell, P. G. Kwiat, J. C. Bienfang, R. P. Mirin, E. Knill, and S. W. Nam, Phys. Rev. Lett. **115**, 250402 (2015).

[5]  C.-Z. Peng, T. Yang, X.-H. Bao, J. Zhang, X.-M. Jin, F.-Y. Feng, B. Yang, J. Yang, J. Yin, Q. Zhang, N. Li, B.-L. Tian, and J.-W. Pan, Phys. Rev. Lett. **94**, 150501 (2005).

[6]  R. Ursin, F. Tiefenbacher, T. Schmitt-Manderbach, H. Weier, T. Scheidl, M. Lindenthal, B. Blauensteiner, T. Jennewein, J. Perdigues, P. Trojek, B. Ömer, M. Fürst, M. Meyenburg, J. Rarity, Z. Sodnik, C. Barbieri, H. Weinfurter, and A. Zeilinger, Nature Physics **3**, 481 (2007).

[7]  S.-K. Liao, W.-Q. Cai, W.-Y. Liu, L. Zhang, Y. Li, J.-G. Ren, J. Yin, Q. Shen, Y. Cao, Z.-P. Li, F.-Z. Li, X.-W. Chen, L.-H. Sun, J.-J. Jia, J.-C. Wu, X.-J. Jiang, J.-F. Wang, Y.-M. Huang, Q. Wang, Y.-L. Zhou, L. Deng, T. Xi, L. Ma, T. Hu, Q. Zhang, Y.-A. Chen, N.-L. Liu, X.-B. Wang, Z.-C. Zhu, C.-Y. Lu, R. Shu, C.-Z. Peng, J.-Y. Wang, and J.-W. Pan, Nature **549**, 43 (2017).

[8]  K. Mattle, H. Weinfurter, P. G. Kwiat, and A. Zeilinger, Phys. Rev. Lett. **76**, 4656 (1996).

[9]  J.-W. Pan, Z.-B. Chen, C.-Y. Lu, H. Weinfurter, A. Zeilinger, and M. Żukowski, Rev. Mod. Phys. **84**, 777 (2012).

[10] D. Bouwmeester, J.-W. Pan, M. Daniell, H. Weinfurter, and A. Zeilinger, Phys. Rev. Lett. **82**, 1345 (1999).

[11] J.-W. Pan, M. Daniell, S. Gasparoni, G. Weihs, and A. Zeilinger, Phys. Rev. Lett. **86**, 4435 (2001).

[12] Z. Zhao, Y.-A. Chen, A.-N. Zhang, T. Yang, H. J. Briegel, and J.-W. Pan, Nature **430**, 54 (2004).

[13] C.-Y. Lu, X.-Q. Zhou, O. Gühne, W.-B. Gao, J. Zhang, Z.-S. Yuan, A. Goebel, T. Yang, and J.-W. Pan, Nature Physics **3**, 91 (2007).

[14] X.-C. Yao, T.-X. Wang, P. Xu, H. Lu, G.-S. Pan, X.-H. Bao, C.-Z. Peng, C.-Y. Lu, Y.-A. Chen, and J.-W. Pan, Nature Photonics **6**, 225 (2012).

[15] X.-L. Wang, L.-K. Chen, W. Li, H.-L. Huang, C. Liu, C. Chen, Y.-H. Luo, Z.-E.

Su, D. Wu, Z.-D. Li, H. Lu, Y. Hu, X. Jiang, C.-Z. Peng, L. Li, N.-L. Liu, Y.-A. Chen, C.-Y. Lu, and J.-W. Pan, Phys. Rev. Lett. **117**, 210502 (2016).

[16] L.-K. Chen, Z.-D. Li, X.-C. Yao, M. Huang, W. Li, H. Lu, X. Yuan, Y.-B. Zhang, X. Jiang, C.-Z. Peng, and others, Optica **4**, 77 (2017).

[17] X.-L. Wang, Y.-H. Luo, H.-L. Huang, M.-C. Chen, Z.-E. Su, C. Liu, C. Chen, W. Li, Y.-Q. Fang, X. Jiang, J. Zhang, L. Li, N.-L. Liu, C.-Y. Lu, and J.-W. Pan, Phys. Rev. Lett. **120**, 260502 (2018).

[18] J.-W. Pan, D. Bouwmeester, M. Daniell, H. Weinfurter, and A. Zeilinger, Nature **403**, 515 (2000).

[19] D. Bouwmeester, J.-W. Pan, K. Mattle, M. Eibl, H. Weinfurter, and A. Zeilinger, Nature **390**, 575 (1997).

[20] X.-L. Wang, X.-D. Cai, Z.-E. Su, M.-C. Chen, D. Wu, L. Li, N.-L. Liu, C.-Y. Lu, and J.-W. Pan, Nature **518**, 516 (2015).

[21] J.-G. Ren, P. Xu, H.-L. Yong, L. Zhang, S.-K. Liao, J. Yin, W.-Y. Liu, W.-Q. Cai, M. Yang, L. Li, K.-X. Yang, X. Han, Y.-Q. Yao, J. Li, H.-Y. Wu, S. Wan, L. Liu, D.-Q. Liu, Y.-W. Kuang, Z.-P. He, P. Shang, C. Guo, R.-H. Zheng, K. Tian, Z.-C. Zhu, N.-L. Liu, C.-Y. Lu, R. Shu, Y.-A. Chen, C.-Z. Peng, J.-Y. Wang, and J.-W. Pan, Nature **549**, 70 (2017).

[22] A. N. Boto, P. Kok, D. S. Abrams, S. L. Braunstein, C. P. Williams, and J. P. Dowling, Physical Review Letters **85**, 2733 (2000).

[23] P. Walther, J.-W. Pan, M. Aspelmeyer, R. Ursin, S. Gasparoni, and A. Zeilinger, Nature **429**, 158 (2004).

[24] I. Afek, O. Ambar, and Y. Silberberg, Science **328**, 879 (2010).

[25] C.-Y. Lu, W.-B. Gao, O. Gühne, X.-Q. Zhou, Z.-B. Chen, and J.-W. Pan, Phys. Rev. Lett. **102**, 030502 (2009).

[26] J. B. Spring, B. J. Metcalf, P. C. Humphreys, W. S. Kolthammer, X.-M. Jin, M. Barbieri, A. Datta, N. Thomas-Peter, N. K. Langford, D. Kundys, J. C. Gates, B. J. Smith, P. G. R. Smith, and I. A. Walmsley, Science **339**, 798 (2013).

[27] M. A. Broome, A. Fedrizzi, S. Rahimi-Keshari, J. Dove, S. Aaronson, T. C. Ralph, and A. G. White, Science **339**, 794 (2013).

[28] M. Tillmann, B. Dakić, R. Heilmann, S. Nolte, A. Szameit, and P. Walther, Nature Photonics **7**, 540 (2013).

[29] A. Crespi, R. Osellame, R. Ramponi, D. J. Brod, E. F. Galvão, N. Spagnolo, C. Vitelli, E. Maiorino, P. Mataloni, and F. Sciarrino, Nature Photonics **7**, 545 (2013).

[30] H. Wang, Y. He, Y.-H. Li, Z.-E. Su, B. Li, H.-L. Huang, X. Ding, M.-C. Chen, C. Liu, J. Qin, J.-P. Li, Y.-M. He, C. Schneider, M. Kamp, C.-Z. Peng, S. Höfling, C.-Y. Lu, and J.-W. Pan, Nature Photonics **11**, 361 (2017).

[31] A. P. Lund, A. Laing, S. Rahimi-Keshari, T. Rudolph, J. L. O'Brien, and T. C. Ralph, Phys. Rev. Lett. **113**, 100502 (2014).

[32] A. Acín, N. Brunner, N. Gisin, S. Massar, S. Pironio, and V. Scarani, Phys. Rev. Lett. **98**, 230501 (2007).

[33] Nature **557**, 212 (2018).

[34] Y. Cao, Y.-H. Li, W.-J. Zou, Z.-P. Li, Q. Shen, S.-K. Liao, J.-G. Ren, J. Yin, Y.-A. Chen, C.-Z. Peng, and J.-W. Pan, Phys. Rev. Lett. **120**, 140405 (2018).

[35] T. B. Pittman, B. C. Jacobs, and J. D. Franson, Phys. Rev. A **66**, 042303 (2002).

[36] A. L. Migdall, D. Branning, and S. Castelletto, Phys. Rev. A **66**, 053805 (2002).
[37] F. Kaneda and P. G. Kwiat, ArXiv:1803.04803 [Quant-Ph] (2018).
[38] Y.-H. Kim and W. P. Grice, Journal of Modern Optics **49**, 2309 (2002).
[39] Y.-H. Kim, S. P. Kulik, M. V. Chekhova, W. P. Grice, and Y. Shih, Phys. Rev. A **67**, 010301 (2003).
[40] S. Takeuchi, Optics Letters **26**, 843 (2001).
[41] O. Kwon, Y.-W. Cho, and Y.-H. Kim, Phys. Rev. A **78**, 053825 (2008).
[42] P. J. Mosley, J. S. Lundeen, B. J. Smith, P. Wasylczyk, A. B. U'Ren, C. Silberhorn, and I. A. Walmsley, Phys. Rev. Lett. **100**, 133601 (2008).
[43] F. Kaneda, K. Garay-Palmett, A. B. U'Ren, and P. G. Kwiat, Opt. Express, OE **24**, 10733 (2016).
[44] C. Chen, C. Bo, M. Y. Niu, F. Xu, Z. Zhang, J. H. Shapiro, and F. N. C. Wong, Optics Express **25**, 7300 (2017).
[45] W. P. Grice, A. B. U'Ren, and I. A. Walmsley, Physical Review A **64**, 063815 (2001).
[46] See Supplemental Material for a comparison of SPDC entangled photon pair sources; joint spectrum without filters; more details about the optical interferometer and validation.
[47] C. K. Hong, Z. Y. Ou, and L. Mandel, Phys. Rev. Lett. **59**, 2044 (1987).
[48] O. Gühne and G. Tóth, Physics Reports **474**, 1 (2009).
[49] S. Aaronson and A. Arkhipov, THEORY OF COMPUTING **9**, 110 (2013).
[50] M. Bentivegna, N. Spagnolo, C. Vitelli, F. Flamini, N. Viggianiello, L. Latmiral, P. Mataloni, D. J. Brod, E. F. Galvão, A. Crespi, R. Ramponi, R. Osellame, and F. Sciarrino, Science Advances **1**, e1400255 (2015).
[51] H. Wang, W. Li, X. Jiang, Y.-M. He, Y.-H. Li, X. Ding, M.-C. Chen, J. Qin, C.-Z. Peng, C. Schneider, M. Kamp, W.-J. Zhang, H. Li, L.-X. You, Z. Wang, J. P. Dowling, S. Höfling, C.-Y. Lu, and J.-W. Pan, Phys. Rev. Lett. **120**, 230502 (2018).
[52] H. Buhrman, R. Cleve, S. Massar, and R. de Wolf, Rev. Mod. Phys. **82**, 665 (2010).

# Figure captions:

**Figure 1.** The design and performance of our new SPDC entangled-photon source. (a) The interferometric two-photon entanglement source at telecommunication wavelength. The laser beam is split into two *H*-polarized beams by two 775-nm BDs and HWPs and focused on a BBO crystal at two different spot to generate photon pairs via type-II SPDC. Then down-converted beams with the same polarization are recombined into one path. The green and red lines represent *H* and *V* polarization respectively. BD: beam displacer. (b) The measured joint spectrum of the photon pair, indicating the two photons are free of frequency correlations. (c) Hong-Ou-Mandel-type interference of two single photons from two independent SPDC as a function of their time delay, measured without using any lossy, narrowband filtering. (d) The measured heralding efficiency and photon-pair brightness as a function of the pump beam waist used in our experiment.

**Figure 2.** Experimental set-up for generating the 12-photon entanglement. Six SPDC entanglement sources (as shown in Fig. 1a) are pumped by laser pulses with a central wavelength of 775 nm, a bandwidth of 5.5 nm and a repetition of 80 MHz. Dispersion of the laser pulses caused by YVO$_4$ crystals and BBO crystals is pre-compensated by 4 prism pairs. The photons pass through 30-nm bandpass filters to remove the sidebands [46] and detected by 24 superconducting nanowire single-photon detectors with an average efficiency of ~75% at 1550 nm. HWP: half-wave plate; QWP: quarter-wave plate; PBS: polarization beam splitter.

**Figure 3.** Experimental result for the 12-photon GHZ state. (a) The measured 12-photon counts in the *H*/*V* basis. (b) Experimentally extracted expectation values of the observables $M_\theta^{\otimes N} = (\cos\theta\sigma_x + \sin\theta\sigma_y)^{\otimes N}$, which is calculated from registered 12-photon coincidence events in the $(|H\rangle \pm e^{i\theta}|V\rangle)/\sqrt{2}$ basis. Error bars stand for one standard deviation based on binomial distribution statistics.

**Figure 4** Experimental setup for the scattershot boson sampling experiment. Photons are produced from 12 individual SPDC sources which are enfolded into 6 BBO crystals. The idler photons act as triggers to herald counterpart signal photons. The signal photons from same crystal are combined into one path by using the same method illustrated in Fig. 1a and guided to a 12-mode optical interferometer. For the boson sampling, we choose a pump beam waist of 0.8 mm and a two-photon count rate of 0.5 MHz. The interferometer multiplexed with 6 spatial modes and 2 polarization modes is consisted of an optical network, a HWP array in the input side and a QWP and PBS array in the output side. All photons are then filtered by 30 nm filters and fed into superconducting nanowire single photon detectors.

**Figure 5** Experimental results of high-efficiency scattershot boson sampling. (a) The measured similarity and distance for the 3-boson sampling. (b) Extended likelihood ratio test between the experimental data and simulated distinguishable sampler for the three-, four- and five-photon experiments. (c) A comparison of the 3-boson sampling rate with previous experiments using SPDC and quantum dots. Each data points are accompanied by their references.

Figure 1

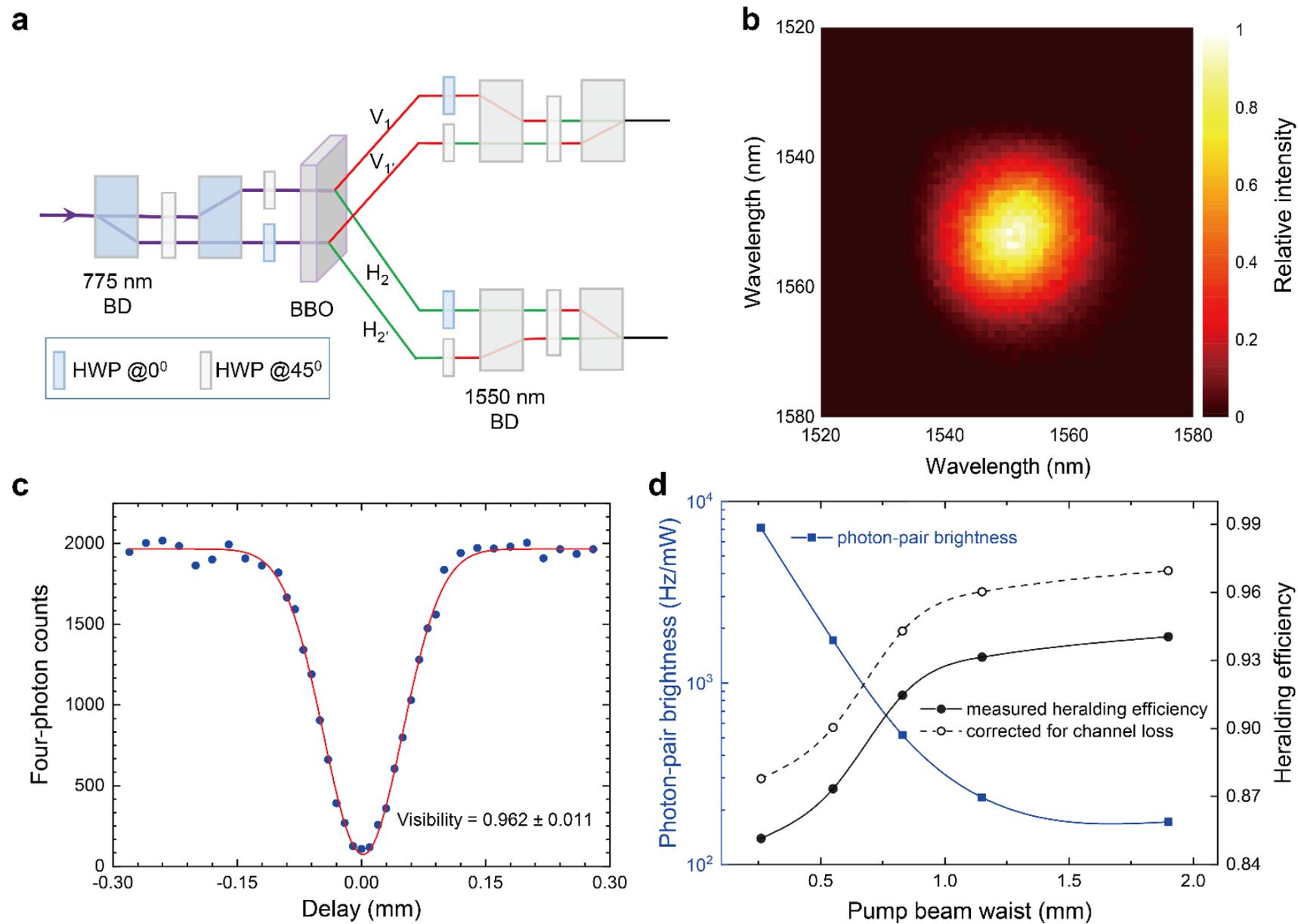

# Figure 2

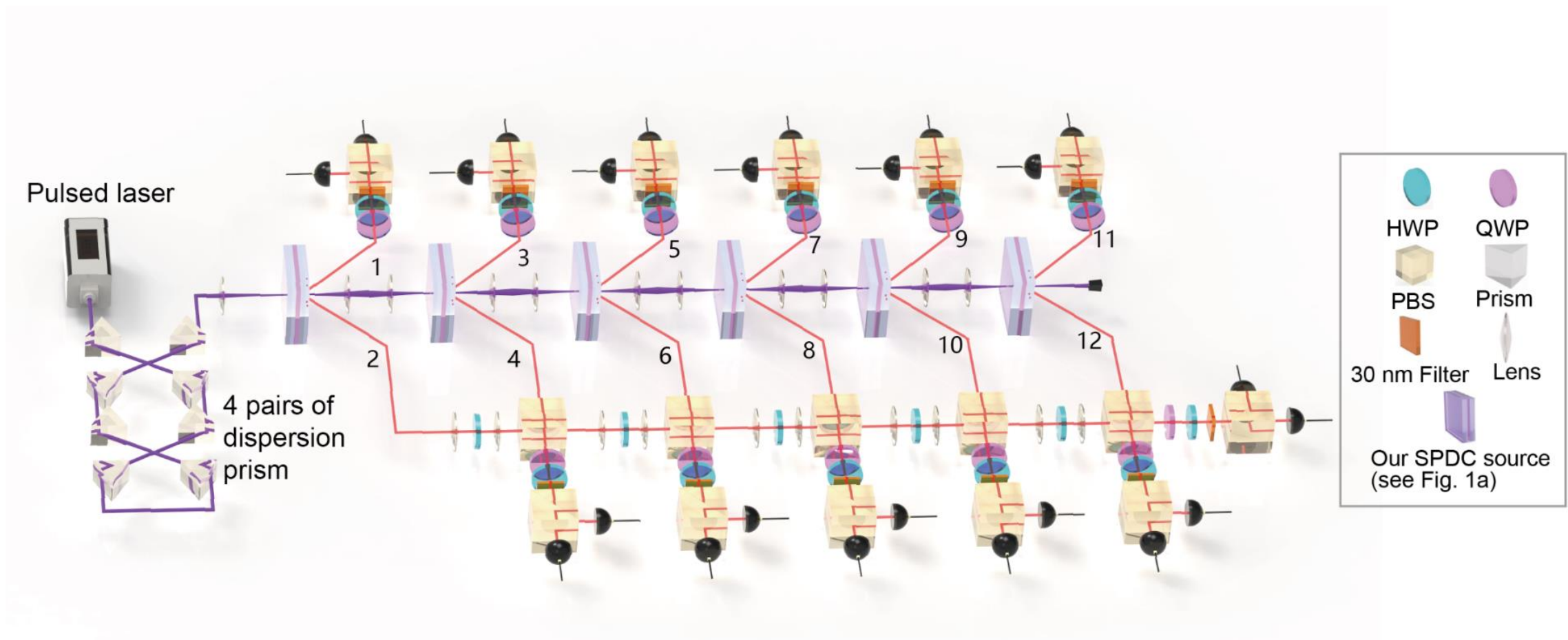

Figure 3

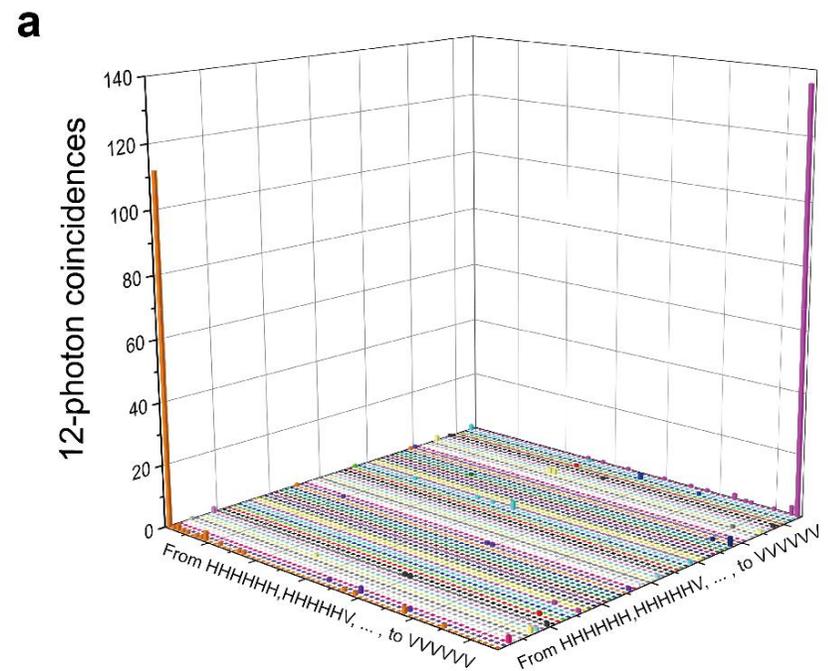

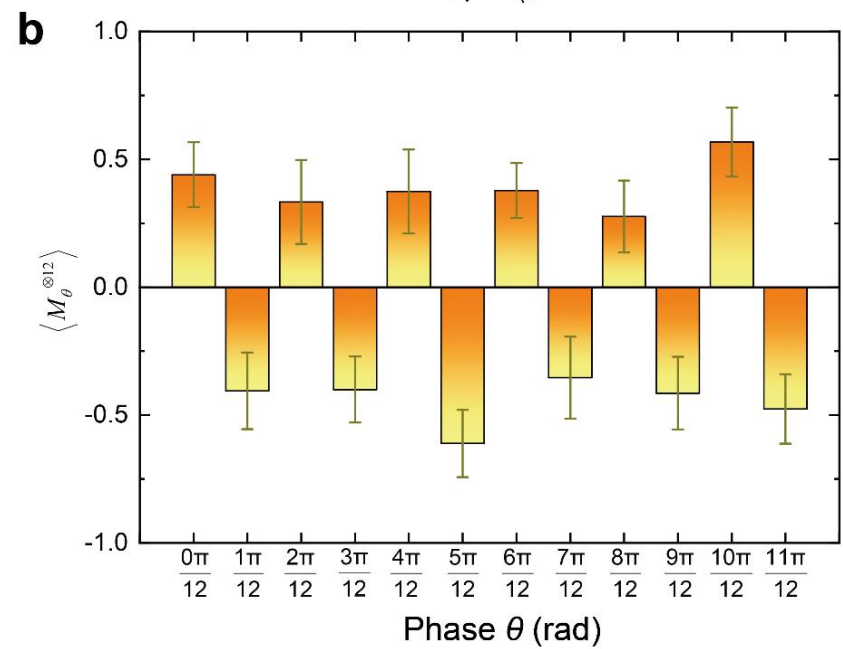

# Figure 4

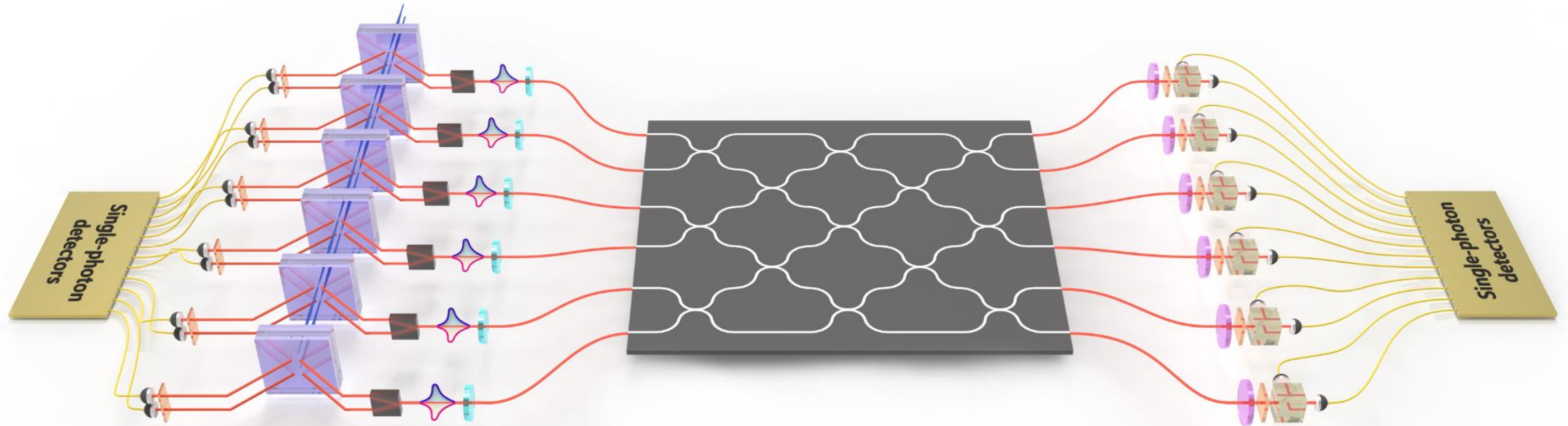

# Figure 5

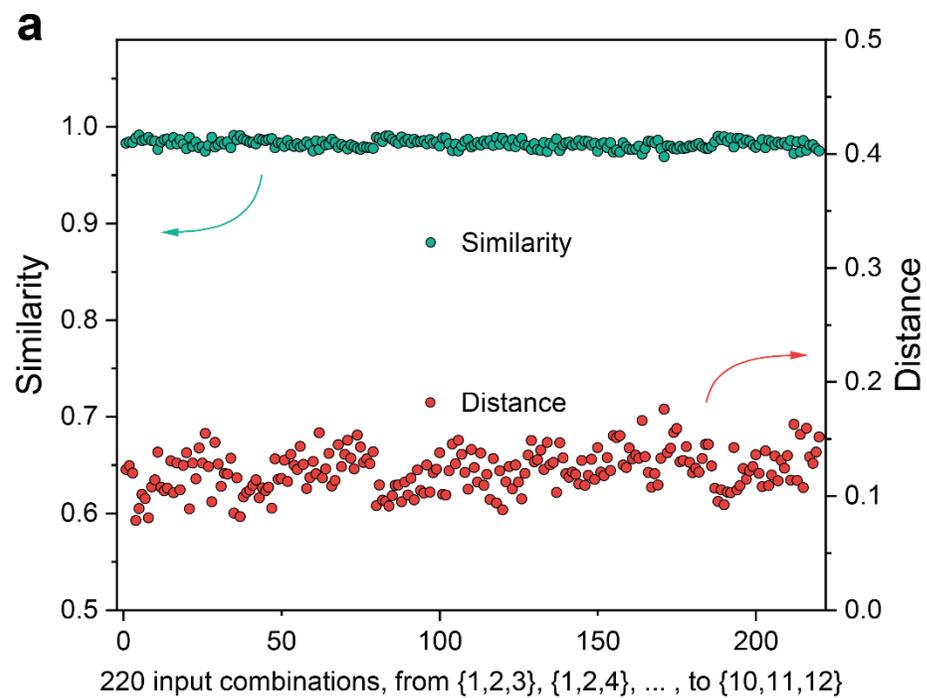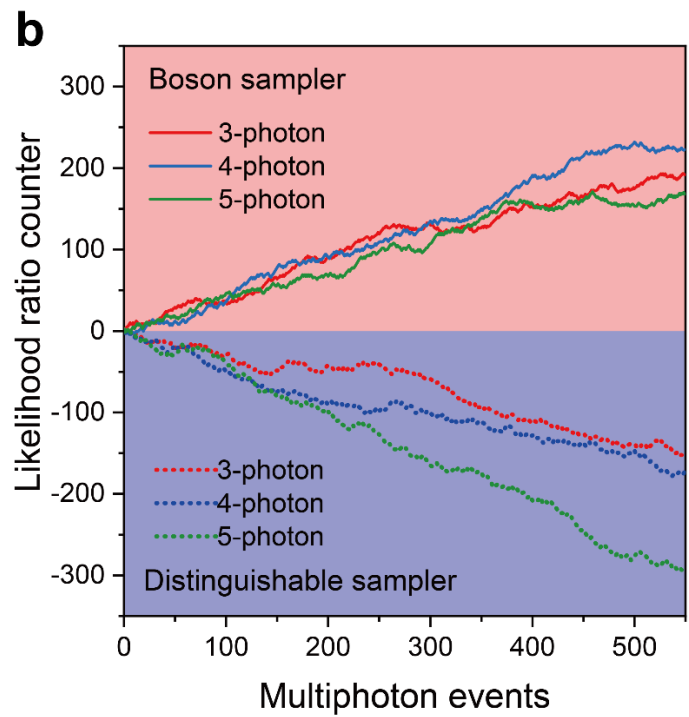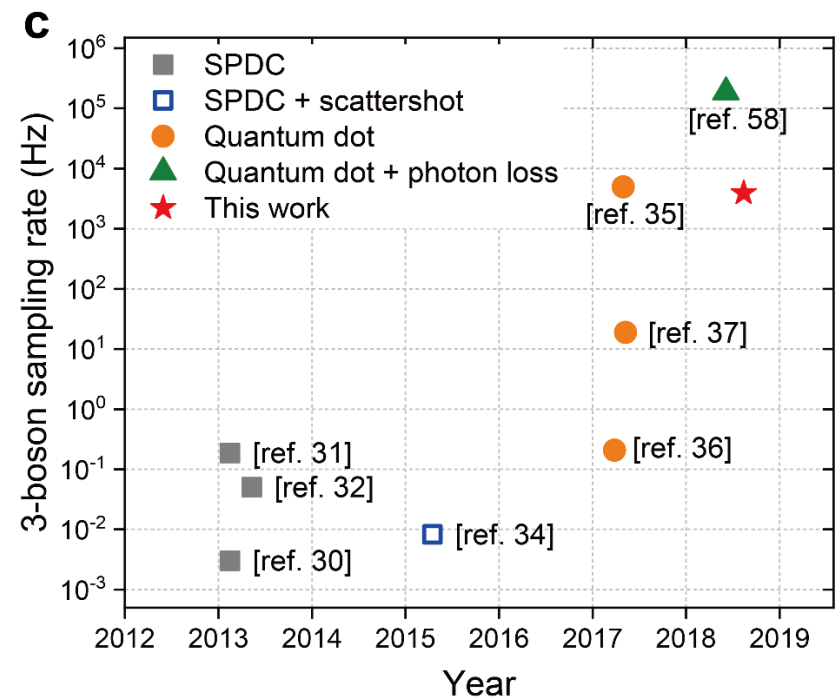